\documentclass[fleqn,usenatbib]{mnras}

\usepackage{newtxtext,newtxmath}
\usepackage{gensymb}
\usepackage{adjustbox,lipsum}
\usepackage{float}
\usepackage{xcolor}
\usepackage[T1]{fontenc}
\usepackage{ae,aecompl}
\defcitealias{matteucci88}{MV88}
\defcitealias{43planck16}{PlanckXLIII-16}
\defcitealias{gutierrez17}{GLC17}
\defcitealias{gutierrez14}{GLC14}

\usepackage{float}
\usepackage{graphicx}	
\usepackage{amsmath}	
\usepackage{amssymb}	

\newcommand*\diff{\mathop{}\!\mathrm{d}}
\newcommand*\e{\mathop{}\!\mathrm{e}}

\title[Galaxy Cluster Dust at low/intermediate z]{On the Origin of Dust in Galaxy Clusters at Low to Intermediate Redshift.} 

\author[E. Gjergo et al.]{
Eda Gjergo,$^{1,2,3}$\thanks{E-mail: gjergo@oats.inaf.it}
Marco Palla,$^{2,4}$
Francesca Matteucci,$^{2,3,5}$ 
Elena Lacchin,$^{2}$ \\~\\
{\rm \LARGE Andrea Biviano,$^{3,4}$  and Xilong Fan.$^{1}$
}
\\ \\
$^{1}$School of Physics and Technology, Wuhan University, Wuhan 430072, China.\\
$^{2}$Dipartimento di Fisica, Sezione di Astronomia, Universit\`{a} di Trieste, Via G.B. Tiepolo 11, I-34131 Trieste, Italy.\\
$^{3}$INAF - Osservatorio Astronomico di Trieste, Via G.B. Tiepolo 11, I-34131 Trieste, Italy\\
$^{4}$IFPU - Institute for Fundamental Physics of the Universe, Via Beirut 2, I-34014, Trieste, Italy\\
$^{5}$INFN - Sezione di Trieste, via Valerio 2, I-34134 Trieste, Italy
}

\date{Accepted XXX. Received YYY; in original form ZZZ}

\pubyear{2019}

\begin{document}
\label{firstpage}
\pagerange{\pageref{firstpage}--\pageref{lastpage}}
\maketitle

\begin{abstract} 
Stacked analyses of galaxy clusters at low-to-intermediate redshift show signatures attributable to dust, but the origin of this dust is uncertain. We test the hypothesis that the bulk of cluster dust derives from galaxy ejecta. To do so, we employ dust abundances obtained from detailed chemical evolution models of galaxies. We integrate the dust abundances over cluster luminosity functions (one-slope and two-slope Schechter functions). We consider both a hierarchical scenario of galaxy formation and an independent evolution of the three main galactic morphologies: elliptical/S0, spiral and irregular. We separate the dust residing within galaxies from the dust ejected in the intracluster medium. To the latter, we apply thermal sputtering. The model results are compared to low-to-intermediate redshift observations of dust masses. We find that in any of the considered scenarios, elliptical/S0 galaxies contribute negligibly to the present-time intracluster dust, despite producing the majority of gas-phase metals in galaxy clusters. Spiral galaxies, instead, provide both the bulk of the spatially-unresolved dust and of the dust ejected into the intracluster medium. The total dust-to-gas mass ratio in galaxy clusters amounts to $10^{-4}$, while the intracluster medium dust-to-gas mass ratio amounts to $10^{-6}$ at most. These dust abundances are consistent with the estimates of cluster observations at $0.2 < z <1$. We propose that galactic sources, spiral galaxies in particular, are the major contributors to the cluster dust budget.
\end{abstract}




\section{Introduction} 

\begin{table*}
    \centering
    \noindent\adjustbox{max width=\textwidth}{%
    \begin{tabular}{l r l r c l  } \hline
    Gal. Clus. Obs. Paper$^{\mbox{(1)}}$ & DtG$^{\mbox{(2)}}$ & $r_{center}^{\mbox{(3)}}$   & $z$ range / cluster$^{\mbox{(4)}}$ & Wavelength (instrument)$^{\mbox{(5)}}$ & Method$^{\mbox{(6)}}$ \\ \hline 
     { \citetalias{gutierrez17}} & $< 9.5 \times 10^{-6}$  (ICM) &  1-5'  & $0.06 < z < 0.7$ & 250, 350, 500 $\mu$m (Herschel) & Stacked em. + Bkg. ext. \\ 
     { \citetalias{43planck16}} & $1.93^{+.92}_{-.92} 10^{-4}$ (Full) & 15'  & $0.01 < z < 1.$  & 850-60 $\mu$m (IRAS/Planck) & Stacked emission \\
     \citetalias{gutierrez14} & $\lesssim 2 \times 10^{-5}$ (ICM) & 3 Mpc & $0.05 < z < 0.68$ & g-r-i (SDSS-DR9)  & Stacked em. + Bkg. ext \\
     \citet{mcgee10} & $\sim 3 \times 10^{-4}$  (ICM)&  $\lesssim 43$ Mpc  & $0.1 < z < 0.2$ & g-r-i-z/12-100 $\mu$m (SDSS/IRAS)   & Background extinction\\
     \citet{roncarelli10} & $\lesssim 5 \times 10^{-5}$ (ICM) &  $< 12$'  &  $0.1 < z < 0.3$ & u-g-r-i-z (SDSS-maxBCG) & SED-reconstruction \\
     \citet{kitayama09} & $< 10^{-5}$ (ICM) & 0.1 Mpc & (Coma cluster) & 24, 70, 160 $\mu$m (Spitzer) & MIR/FIR emission \\
     { \citet{bovy08}} & $\lesssim 2 \times 10^{-5}$ (ICM) & 2 Mpc & $z \sim 0.05$ & u-g-r-i-z (SDSS-DR4) & Background extinction \\
     \citet{giard08} & $\lesssim 10^{-5}$ (ICM) & 10'  & $0.01 < z < 1$ & 12-100$\mu$m/0.1-2.4keV (IRAS/RASS) & Stacked emission \\
     \citet{muller08} & $ \lesssim 2 \times 10^{-4}$ (ICM) & 1.5 Mpc & $z < 0.5$ & 650, 910 $\mu$m (CFHT)  & Background extinction \\ 
     \citet{chelouche07} & $< 5 \times 10^{-4}$ (ICM) & $\sim 1$ Mpc  & $0.1 < z < 0.3$ & u-g-r-i-z (SDSS)  & Background extinction\\
     { \citet{stickel02}} & $\sim 10^{-6}$ (ICM) & 0.2 Mpc &  (Coma cluster+)  & 120, 180 $\mu$m (ISO) & $I_{120}/I_{180}$  
     \\
     \hline 
    \end{tabular}}
    \caption{$^{\mbox{(1)}}$For various observational papers of galaxy clusters  -- where the aliases \citetalias{gutierrez14, gutierrez17, 43planck16} refer to \citet{gutierrez14, gutierrez17, 43planck16} respectively -- $^{\mbox{(2)}}$the Dust-to-Gas (DtG) ratio estimates. The parentheses (Full) and (ICM) differentiate respectively the DtG estimates over the total cluster from the estimates that isolate the ICM component. $^{\mbox{(3)}}$The radius --- in arcminutes or Megaparsecs --- within which the estimate is computed, $^{\mbox{(4)}}$the redshift spanned by the study or by the observed cluster, $^{\mbox{(5)}}$the instrument used to gather data and the relative observed wavelengths or bands, and $^{\mbox{(6)}}$the method employed to extrapolate the dust estimate. Specifically, $I_{120}/I_{180}$ ia the surface brightness ratios in the FIR; Background extinction refers to the reddening of galaxies and quasars located in the background of galaxy clusters.  }
    \label{tab:galclusobs}
\end{table*}

Even at low abundances, dust is an important ingredient in both the observation and the evolution of astrophysical systems. Dust affects the spectral energy distributions (SED) of any target, particularly stars and galaxies, from the UV-and-optical to the far-infrared  \citep[FIR, e.g., ][]{calzetti01}. 
It is estimated that about 50\% of all the starlight emitted across cosmic history has been reprocessed by dust \citep{hauser01}. 
{ Dust depletes galactic interstellar media (ISM) of about half of their gaseous metals, many metals which (e.g., O, C, Fe) are important coolants.  On the other hand, dust serves as a catalyst for the formation of some of the most important cooling molecules, especially H$_2$ \citep{gould63}.  These dust-related phenomena affecting gas temperature could indirectly influence star formation (SF). }
 
Despite many open issues { regarding the composition and size distribution of dust \citep[e.g., ][]{jones14}, the field has made substantial progress in the understanding of dust properties within galaxies \citep[for a review, see ][]{galliano18}}. However, much less is known regarding the abundance, and size distribution of dust in galaxy clusters -- including their diffuse intracluster media (ICM).
The understanding of dust distribution in galaxy clusters is important for a number of reasons. For example, it affects galaxy cluster catalogue completeness. Up to $\sim $ 9\% of clusters in a redshift range of $0.5 < z < 0.8$ might be undetected in the Planck survey due to the presence of dust \citep{melin18}. Dust becomes an efficient cooling agent in the hot temperatures ($T > 10^7$K) typical of the ICM \citep{montier04}. Furthermore, a proper understanding of dust properties will be essential in the analysis of protoclusters at redshift $z \gtrsim 2$ \citep[e.g.,][]{casey16, cheng19, smith19}. These systems are not fully virialized, so their ICM has not reached the temperatures necessary to trigger efficient destruction mechanisms (i.e., sputtering).

There is no definitive evidence of the presence of dust in the ICM of local galaxy clusters. \citet{stickel02} put forward an estimate of a dust-to-gas mass ratio (DtG) for the ICM of the Coma cluster of about $10^{-6}$; it however found no evidence of dust in other clusters (A262, A2670, A400, A496, A4038). Later \citet{kitayama09}, while finding estimates consistent with \citet{stickel02} for the Coma cluster DtG, attributed the low dust abundances to irregular sources fluctuating in the cirrus foreground. Similarly, \citet{bai07} found that the emission from the Abell cluster A2029 is indistinguishable from cirrus noise.

Evidence of the presence of dust in clusters may instead  have been found in the statistical analysis of large datasets. \citet{chelouche07} and \citet{muller08}, in the redshift ranges $0.1 < z < 0.3$ and $z < 0.5$ respectively, observed dust extinction in galaxies and quasars located in the background of galaxy clusters. Both obtain comparable DtG measures of $ 10^{-4}$. \citet{chelouche07} estimated a DtG smaller than $< 5$\% of the typical galactic ISM values; \citet{muller08} found a dust mass upper limit of $8 \times 10^9 M_{\odot}$ within $R_{200}$\footnote{$R_{200}$ ($M_{200}$) is the radius that encloses (mass enclosed by) a sphere whose mean density is 200 times the critical density at the given redshift.}.
{ \citet{chelouche07} and later \citet{gutierrez17} noted that the ICM dust is mostly distributed in the outskirts of clusters. The dust signature drops closer to the cluster center, in the very hot ICM.}  This same radial dependence was observed by \citet{mcgee10} { in systems within a redshift range of $0.1 < z < 0.2$, out to} larger radii of 30$h^{-1}$Mpc from the center of large clusters ($\sim 10^{14} h^{-1} M_{\odot}$) and small groups ($\sim 10^{12.7} h^{-1}M_{\odot}$).

Analysing IR data, \citet{giard08} found more stringent upper limits. 
{ On a selection of galaxy clusters from three catalogues} \citep{gal03, montier05, koester07}, they stacked the integrated IR  luminosity \citep[$L_{IR}$, ][]{miville05} within an annulus between 9' and 18', in redshift bins up to $z < 1$.
They then paired the IR data with the X-ray Rosat All Sky Survey (RASS) \citep{voges99} for all of the selected clusters, enabling the identification of the ongoing SF from member galaxies. After subtracting the SF component from the estimates, the resulting signal attributable to ICM dust leads to a DtG not greater than $10^{-5}$.

\citet{roncarelli10} is a follow-up to \citet{giard08} on a restricted redshift range of $0.1 < z < 0.3$ that employs the SDSS-maxBCG  catalogue \citep{koester07}.  {The galaxies in this calalogue were separated by morphology -- namely  E/S0, Sa, Sb, Sc and starburst -- and for each, they modeled their respective SEDs. They then reconstructed the 60 and 100 $\mu$m IRAS band emissions of the cluster galaxies using the prescriptions by \citet{silva98}. Finally, they compared their predictions to the IR data by \citet{giard08} to isolate the IR emission not coming from dust in known galaxies, i.e. the emission coming from ICM dust.} Their estimated total galactic emission is dominated by star-forming late-type galaxies, leading to an estimated  ICM DtG of $\lesssim 5 \times 10^{-5}$.

{ \citet{43planck16}, hereafter \citetalias{43planck16}, and \citet{gutierrez14,gutierrez17}, hereafter \citetalias{gutierrez14} and \citetalias{gutierrez17}, provided} some of the latest estimates { on dust in galaxy clusters}. { \citetalias{43planck16} observed} integrated -- and hence not spatially-resolved -- cluster dust masses of a few $10^{10} M_{\odot}$ within a fixed aperture of 15 arcmin for massive galaxy clusters within a redshift of $z < 1$. The integration was run over a few reasonable values of the spectral emission fitting parameters. It is therefore an estimate of the total dust mass, including both galaxy (ISM) dust and ICM dust. \citetalias{gutierrez14} performed an analysis of both stacked IR emission and of background object extinction in an effort to disentangle the contributions of dust coming from the ICM { and of dust residing within the ISM of} cluster galaxies.
All these studies, amid uncertainties, detected low dust abundances that may impact the interpretation of star formation rates (SFR) and evolutionary models for both galaxies and galaxy clusters.


Sporadic theoretical works have attempted to estimate ICM dust. \citet{dwek90} already predicted that dust should exist { mostly far away from the cluster center ($R > 2$ Mpc) and in low abundances}. A decade later, \citet{popescu00} proposed that any IR emission by diffuse ICM dust would come from current dust injection in the ICM, and hence it would indicate the dynamical state and maturity of the cluster. More recently, \citet{polikarpova17} determined that dust may live in the ICM for 100-300 Myr if it  resides in isolated dense and cold gas filamanets surviving the outflow. This would lead to an ICM DtG of about 1-3\% of the typical Galactic values, whose average DtG is $10^{-2}$. 
Some hydrodynamical simulations and semi-analytical models (SAMs) of galaxies and galaxy clusters have already included dust evolution \citep[e.g., ][]{bekki15, zhukovska16, popping17, aoyama17, mckinnon17, gjergo18, vogelsberger19, hu19}. Among these, \citet{popping17} with SAMs, and \citet{gjergo18} and \citet{vogelsberger19} with cosmological simulations of galaxy clusters and galaxies respectively, treated dust destruction by thermal sputtering in the harsh extragalactic and intracluster environments. \citet{gjergo18} slightly underproduced dust compared to \citetalias{43planck16}, and alleviates the tension by relaxing the sputtering destruction timescale. \citet{vogelsberger19} is able to reproduce the Planck results by also relaxing the sputtering timescale, and by including  gas cooling due to dust in high resolution simulations.

In this work we present an approach previously tested on gas metals for local galaxy clusters by \citet{matteucci88} \citepalias[hereafter][]{matteucci88}. The method consists in integrating at present time the chemical evolution models of elliptical/S0 galaxies over luminosity functions (LF). This approach successfully predicted that the bulk of the metal mass is produced by elliptical/S0 galaxies at the break of the LF \citep{gibson97}. 

We implement the same technique of \citetalias{matteucci88}, but using dust evolution models, and we apply it to the entire evolution history of a typical cluster. The comprehensive dust prescriptions have been validated in the solar neighborhood, damped Lyman alpha systems, far away galaxies and quasars, as well as across cosmic times \citep{fan13, gioannini17a, gioannini17b, spitoni17, vladilo18, palla19, palla19b}.
 Even though these chemical and dust evolution models have been mainly calibrated on field galaxies, it is safe to apply them to cluster galaxies. In fact, some studies \citep[e.g.,][]{davies19dustpedia} have shown that dust properties such as DtG and dust-to-stellar-mass ratios vary more stringently with morphology, age, and physical processes  rather than with environment (field galaxies or cluster galaxies).  

Unlike \citetalias{matteucci88}, we differentiate among the three main morphologies: elliptical/S0, spiral and dwarf/irregular galaxies. For each of these, we separate the dust component residing within the ISM of galaxies from the other component ejected in the ICM.
On top of the standard Schechter function \citep{schechter76}, we test the behavior of a double LF that consists of the sum of two LFs -- one for massive galaxies, and one for dwarf/irregular galaxies. The parameters for both single and double LF follow \citet{moretti15}, that derived median and average best fits for the full sample of the WINGS  low redshift clusters. WINGS \citep[WIde–field Nearby Galaxy–cluster Survey;  ][]{fasano06, moretti14}  is an all-sky survey of nearby galaxy clusters. WINGS comprises all clusters from three  X-ray flux-limited samples in the redshift range 0.04-0.07, and with a Galactic latitude   $\mid b \mid > 20^{\circ}$. The completeness of the survey and the availability of relatively  deep photometry in the B and V band, makes the WINGS sample ideal for the analysis in this paper.

Aside from the scenario where different galaxies evolve independently (monolithic open-box scenario), we expand the model to account for a hierarchical galaxy formation scenario following the method presented in \citet{vincoletto12}. In this model, the galaxy number density $n^*_k$ for different morphologies was tuned to reproduce two quantities: the galaxy fractions at present time ($z < 0.05$) \citep[][]{marzke98}; and the cosmic star formation rate predicted in \citet{menci04} by means of a semi-analytical hierarchical model. This same method was adopted and applied to dust estimates in \citet{gioannini17b}. 

The paper is organized as follows: in Section \ref{sec:obsdata} we overview the relevant observational papers that investigated the presence of dust in galaxy clusters. In Section \ref{sec:methods} we describe in detail the methodology employed, including a summary of the the integration method and dust evolution models. 
In Section \ref{sec:results} we present our predictions of the dust evolution within a typical galaxy cluster, and we compare it against the latest observations.
We also present the dependence of dust evolution in clusters on a reasonable range of parameter values. Finally, our discussion and conclusions are accounted for in Section \ref{sec:conclusions}.

{ Throughout this paper, we adopt a flat $\Lambda$CDM cosmology with a Hubble constant of $H_0 = 70$ km s$^{-1}$ Mpc$^{-1}$ and $\Omega_m = 0.3$.}

\section{Observations of Dust in Galaxy Clusters}\label{sec:obsdata}

A summary of the existing observational literature is presented in Table \ref{tab:galclusobs}. In bold are the observational { studies that we compared to our results}.
Some works investigated individual clusters \citep[e.g., ][]{stickel98, stickel02, bai07, kitayama09} and some took a statistical average over large data sets { investigating optical extinctions} \citep[][]{chelouche07, muller08, mcgee10} or dust IR emission \citep[][]{giard08, gutierrez14, gutierrez17, 43planck16}. Lastly, \citet{roncarelli10} reconstructed the SED of various galactic morphologies using both SDSS and IRAS data, in order to isolate a galactic SED signal from the ICM dust.

It is possible to estimate dust abundances either through dust emission in the IR or through the extinction { of objects on the background of the observed medium} in UV-optical wavebands. Typically, the IR dust emission technique consists in fitting the IR fluxes on modified blackbody spectra of dust thermal emission \citep[e.g.,][]{hildebrand83}. 
A technique for ICM extinction was pioneered by \citet{ostriker84}. 
They estimated dust extinction in a given cluster by measuring the flux of objects -- galaxies and quasars -- located in the background of the given cluster. The dust-obscured flux is then compared to a reference flux of similar objects located at a similar redshift, but in the field, away from clusters { and dust contamination}. Employing this method, \citet{ferguson93} and \citet{maoz95} found that whatever dust may be contained in the ICM of galaxy clusters, it should be negligible compared to selection effects. 

In general, dust abundances in the ICM of large cluster samples at redshift $z < 1$ are not too well constrained, but most dust estimates limit the ICM DtG to around $10^{-5}$, which is around 3 orders of magnitude lower than the typical Galactic ISM values. Such low abundances are due to the short dust destruction timescales in the hostile ICM environment, which is permeated with X-ray radiation and highly energetic ions. Therefore, ICM dust is believed to be of recent origin \citep[e.g., ][]{dwek90, popescu00, clemens10} -- it is either newly ejected from galaxies by stellar winds, or stripped from the galactic ISM by merging events and ram pressure stripping.
Dust is furthermore expected to reside mainly in the outskirts of the cluster, where late-type galaxies are dominant, and where the environment is contaminated by small groups or residues of past mergerer events. This is corroborated by cluster dust profile studies \citep{chelouche07, muller08} in combination with the low dust abundances observed around cluster centers \citep{stickel02, bai07, kitayama09}.

In our work, we compare the obtained results with data by \citet{stickel02}; \citetalias{gutierrez17, 43planck16}.

\citet{stickel02} predict the Coma cluster ICM dust. We take their value as the upper limit for dust content in the ICM of local galaxy clusters.

\citetalias{gutierrez14} employed two methods on the SDSS-DR9 \citep{ahn12} sample of galaxy clusters{  located in a redshift range of} $0.05 < z < 0.68$: the first method is a statistical approach to extinction. Their prediction for total dust mass averages  $M_{d} < 8.4 \times 10^9 M_{\odot}$ within a cluster radius of 3 Mpc. 
The second method 
 is an emission estimate of the contribution of galaxy cluster dust to the FIR sky from optical extinction maps
. This second method leads to a lower prediction of $\sim 2 \times 10^9 M_{\odot}$. 
The conservative DtG upper limit from the two methods combined is $\sim 8 \times 10^{-5}$.
Later, \citetalias{gutierrez14} was followed up through the \emph{Herschel HerMES} project by \citetalias{gutierrez17}. The cluster selection sample contained 327 clusters. \citetalias{gutierrez17} { binned the estimates in three redshift bins (0.05--0.24, 0.24--0.42, 0.41--0.71), two cluster mass bins ($<10^{14} M_{\odot}$ and $> 10^{14} M_{\odot}$), aperture (1 to 5 arcmin), and observed frequency (250, 350, and 500 $\mu$m)}. Our theoretical predictions are compared to a selected sample of \citetalias{gutierrez17} data, in particular to the three redshift bins for the massive cluster sample measured through the 350 $\mu$m channel, for arcmin 1'.

\citetalias{43planck16} considered a selection of 645 clusters within a redshift of $z < 1$. For these clusters, they combined the Planck-HFI maps (6 beams, 100 to 857 GHz) with the IRAS \citep{miville05} maps (60 and 100 $\mu$m), 
they then integrated the stacked signal for each beam out to an aperture of 15 arcmin. Fixing the aperture radius implies that for more distant clusters, more of their outskirts is included in the analysis. They hence fit these 7 data points to the IR SED dust emission, following the approach prescribed in \citet{hildebrand83}.   \citetalias{43planck16} ignores IRAS data at 60$\mu$m in the SED fit, because at this wavelength the contribution by small grains which are not in thermal equilibrium becomes prominent { and would skew the fit}.
For the full sample, each cluster is estimated to have, within 15 arcmin, a dust mass of around $10^{10} M_{\odot}$ -- with small variations depending on the choice of emissivity index $\beta$. The full sample {  ($\langle z \rangle = 0.26 \pm0.17$, $M_{dust} = 1.08 \pm0.32 \times 10^{10} M_{\odot}$)} is split in two redshift bins and two mass bins. The redshift bins are divided at $z = 0.25$, with an average dust mass of $0.34 \pm 0.17 \times 10^{10} M_{\odot}$ for the low $z$ and $2.56 \pm 0.91 \times 10^{10} M_{\odot}$ for the intermediate $z$. The mass bins are divided at $M_{200} = 5.5 \times 10^{14} M_{\odot}$. In this case, the less massive clusters have on average a dust mass of $0.21 \pm 0.14 \times 10^{10} M_{\odot}$ and the more massive clusters fair at $3.48 \pm 0.99 \times 10^{10} M_{\odot}$. Comparisons are made both with the full sample and the two subsample{ s} split according to redshift bins. 


\section{Method} \label{sec:methods}
\sloppy
In order to compute the total amount of dust produced and ejected by galaxies in the ICM, we follow the method proposed in \citetalias{matteucci88}, { in which they integrated over the LF of clusters the masses of given chemical species $i$ produced by galaxies at the present time. Unlike \citetalias{matteucci88} which employs only chemical evolution models of early-type galaxies \citep{matteucci87}, we take advantage of dust evolution models of irregular, spiral, and elliptical galaxies \citep[i.e.][]{gioannini17a, palla19}. We also extend \citetalias{matteucci88} across the entire evolutionary history of the cluster. To do so, we assume two scenarios of galaxy formation: in one case, we consider a monolithic evolution of the cluster, where the three galaxy morphologies evolve indipendently, not changing in their number and abiding only by the chemical and dust evolution models. In the second case, we adopt the \citet{vincoletto12} prescription for hierarchical clustering, already paired to our dust evolution models in \citet{gioannini17b}. What follows is the presentation of  the monolithic scenario. We will then introduce the hierarchical scenario in Section \ref{sec:hierarchical} and the chemical and dust evolution models in Section \ref{sec:dustmodels}. } 

\subsection{Modeling dust in galaxy clusters} \label{sec:modelingbasic}

{ First, we find the relationship between the evolution of dust masses $M_d$ across the cosmic time $t$ for galaxies of a set morphology and total 
baryonic galaxy mass $M_{G}$. This quantity includes gas infall and outflow, and in the rest of the paper we will refer to it as "infall mass". We iterate our dust evolution code over a range of masses $M_{G,X}$ for each of the three morphologies $X$. In the case of elliptical galaxies, $M_{G,ell}$ has a lower limit of $10^9 M_{\odot}$ and does not extend beyond $10^{12} M_{\odot}$. 
 $5 \times 10^9 M_{\odot} < M_{G,spi} < 5 \times 10^{11} M_{\odot}$ is  the range for spiral galaxies, and $10^7 M_{\odot} < M_{G,irr} <  5 \times 10^9 M_{\odot}$ for irregular galaxies. For each iteration, we are able to separate an "ISM" dust component, residing within galaxies, and an "ICM" dust component, ejected by stellar winds.

For each morphology across the respective iterations of dust evolution, we fit the following: }

\begin{equation}\label{eq:speciestogalmass}
    M_d \left(t\right) = E_d\left(t\right) M_{G,X}\,^{\beta_d\left(t\right)},
\end{equation}

\noindent { where $E_d(t)$ and $\beta_d(t)$ are the time-dependent fit parameters. The fits are stable across cosmic time for every iteration.}

{ With the relation between dust and galaxy mass established, we convert galaxy masses into luminosities, because our ultimate goal is to use Equation \ref{eq:speciestogalmass} as a weight function on the LF. We consider the mass-to-light ratio $K = M_G/L$, where the galaxy mass $M_G$ and luminosity $L$ are both expressed in solar units. We take a fiducial value of $K=5$, and test variations up to $K=15$. This range of $K$ is typical of several studies in literature  \citep[e.g.,][]{spiniello12,demasi19,portinari04} for both early and late-type galaxies. 
It is then possible to normalize Equation \ref{eq:speciestogalmass} by the respective quantities at the break ($^*$) of the LF: 
\begin{equation}\label{eq:norm}
M_d / M_d^* = \left(M_G / M_G^*\right)^{\beta_d} = \left(L / L^*\right)^{\beta_d},
\end{equation}

\noindent where $M_d^*$ is the dust mass associated to a galaxy of mass $M_G^*$ and luminosity $L^*$ at the break.
 
 With these tools, we can now consider the distribution function of galaxies across luminosity (or mass). The most reasonable choice is the Schechter LF \citep{schechter76}: $\Phi\left(L\right) = n^* (L/L^*)^{\alpha} \e^{-L/L^*}$, where $L^*$ is the luminosity of a galaxy at the break of the Schechter Function, $n^*$ is a measure of the cluster richness (the number of galaxies per unit luminosity $L_{\odot}^{-1}$), and $\alpha$ is the dimensionless slope of the power law. We weight $\Phi\left(L\right)$ by the normalized Equation \ref{eq:norm}. As $\Phi\left(L\right)$ includes all morphologies, we need to rescale the integration by the number fraction $f_X$ of each morphology $X$ compared to the total number of galaxies. In the Coma Cluster, for example, $f_{ell} = 0.82$. In dynamically young clusters such as Virgo, the elliptical fraction has a lower value of $f_{ell} = 0.43$. The spiral number fraction is then $f_{spi} = 1 - f_{ell}$. We will define the irregular fraction in the next paragraph.
 
The integrand that yields the cluster dust mass for a given morphology then takes the form of 
\begin{equation}
(L/L^*)^{\beta_{d,X}} \Phi\left(L\right) = f_X n^* (L/L^*)^{\beta_{d,X} + \alpha}\e^{-L/L^*},
\end{equation}
 
\noindent which can be integrated as an upper incomplete Gamma function $\Gamma(a,x) = \int_{x}^{\infty}t^{a-1}\e^{-t} \diff t$, where $x$ is the lower limit of the integral. This is appropriate in the case of elliptical galaxies. For spiral and irregular galaxies we impose an upper limit to the integration, subtracting a second incomplete Gamma function to crop out the regions of the LF where we do not observe this morphology. We will see that in the case of spiral galaxies, this narrower range does not affect significantly the dust mass. To ensure we do not count galaxies in multiple morphologies, we impose the same value for the upper bound of irregular galaxies and the lower bounds of spiral and elliptical galaxies, and we take $f_{irr} = 1$. 

The cluster dust mass for each morphology $X$ within the ISM (and similarly for the ejected ICM component) is then derived as:

\begin{multline}\label{eq:SchechterIntegral}
    M_{d,ell}^{ISM} (< R_{200}) =  f_{ell} M_{d,ell}^{*,ISM} n^* \Gamma(1 + \beta_{d,ell}^{ISM} + \alpha, L_{min, ell}/L^*) 
\end{multline}
\begin{multline}
    M_{d,spi}^{ISM} (< R_{200}) =  f_{spi} M_{d,spi}^{*,ISM} n^* \lbrack \Gamma(1 + \beta_{d,spi}^{ISM} + \alpha, L_{min, spi}/L^*)\\
      - \Gamma(1 + \beta_{d,spi}^{ISM} + \alpha, L_{max, spi}/L^*)\rbrack 
\end{multline}
\begin{multline}
    M_{d,irr}^{ISM} (< R_{200}) =  f_{irr} M_{d,irr}^{*,ISM} n^* \lbrack \Gamma(1 + \beta_{d,irr}^{ISM} + \alpha, L_{min, irr}/L^*)  \\
     - \Gamma(1 + \beta_d^{irr,ISM} + \alpha, L_{max, irr}/L^*)\rbrack,
\end{multline}
}
\noindent where $L_{min, X}$ is the lowest luminosity observed in a given cluster and $M_{d,X}^* = E_{d,X}M_G^{*\beta_{d,X}(t)}$ is the dust mass for a galaxy at the break of the LF for the $X$ morphology { in the ISM (or ICM)}. $M_G^*$ is the break galaxy { baryonic} mass.
 
 Observational constraints on magnitude will define our luminosity integration limits, but also our galaxy masses. As derived in \citetalias{matteucci88},
$M^*_G = K L^* = h^2 K * 10^{- 0.4 (\mbox{Mag}^* - \mbox{Mag}_{\odot})}$. $ \mbox{Mag}_{\odot} = 4.83$ and $\mbox{Mag}^*$ are respectively the V-band magnitude of the Sun and of a break galaxy.
 
{ Aside from the original Schechter LF, we also test a double Schechter LF. In the rest of the text, we will refer to them respectively as single LF and double LF. The double LF \citep[e.g.,][]{popesso06}, is in \citet{moretti15} the best fit to the WINGS survey. }
The function has the form of:

\begin{equation}\label{eq:doubleSchechter}
\Phi(L)\propto \Phi_b + \frac{L^*_b}{L^*_f}\Phi_f  =   \left(\frac{L}{L^*_b}\right)^{\alpha_b}\e^{-L/L^*_b} + \frac{L^*_b}{L^*_f} \left(\frac{L}{L^*_f}\right)^{\alpha_f}\e^{-L/L^*_f},
\end{equation}

\noindent where $\Phi_b$ and $\Phi_f$ are single Schechter functions calibrated on the bright end ($b$) and on the faint end ($f$) of the LF. Each of $\Phi_b$ and $\Phi_f$ have their own bright and faint break point, identified with respective break luminosities $L^*_b$ and $L^*_f$ and power law coefficients $\alpha_b$ and $\alpha_f$. We weight spiral and elliptical galaxies with the bright component. { For irregular galaxies, we treat the bright component as we treat irregulars in the single LF, and we add to it the incomplete gamma function integration of the faint component.} { By steepening its slope at fainter luminosities, the double LF} predicts the existence of more dwarf irregular galaxies than the normal one-slope Schechter function. 

The parameters $\alpha$ for the single LF (or $\alpha_b$ and $\alpha_f$ for the double LF) and $L^*$ (or $L^*_b$ and $L^*_f$) are unique for individual clusters, and are taken from the WINGs \citep{moretti15} median parameters unless otherwise specified: $\alpha = -1.15$ (or $\alpha_b = -0.97$ and $\alpha_f = -0.6$ for the double LF). 
The median V-band break magnitude is $\mbox{Mag}_V^* = -21.30$ (or -21.15 and -16.30 for the bright and faint end of the double LF). { $f_{ell} = 0.74$ is the average elliptical fraction for WINGs we extrapolated from \citep{mamon19}. To obtain the cluster richness, we take advantage of the richness-to-cluster-mass relation $n^* \propto M_{clus}^{0.92}$ found by \citep{popesso07} and we rescale the Coma cluster richness to WINGS masses. The Coma cluster has a mass of $M_{clus} \sim 1.1 \times 10^{15} M_{\odot}$ \citep{geller99}, while WINGS clusters average to $M_{clus} \sim 5 \times 10^{14} M_{\odot}$ \citep{mamon19}. $n^*$ for Coma is 107 \citep{schechter76}, it follows from the \citet{popesso07} relation that $n^* = 52$ for the average WINGS cluster sample.} 
We do not report other well-known local clusters to avoid redundancy due to the similarity of the results, { but we present a reasonable range of the parameter space to test what the dust evolution might look like in other clusters}.   We chose to express our integration limits in terms of mass ranges, as defined at the beginning of this section. Note that the irregular galaxy lower integration bound of $1 \times 10^7 M_{\odot}$ corresponds to the WINGs faintest magnitude limit of Mag$_{min} \simeq -15.5$ \citep{moretti15}. We refrain from integrating the LF beyond the WINGS faintest magnitude limit to avoid extrapolations unwarranted by the data, but we note that the contribution of galaxies
 fainter than this limit is not substantial (see Section \ref{sec:fig3}). The parameters just presented are summarized in Table \ref{tab:singleLF}, and in Table \ref{tab:doubleLF} we report the double LF parameters. 

\begin{table}
    \centering
         \begin{tabular}{c c c}
         \hline \hline
        $M_{G,min}^{Irr}$ & $M_{G,min}^{Spi}$ & $M_{G,min}^{Ell}$ \\ \hline
        $10^7 M_{\odot}$ & $5 \times 10^9 M_{\odot}$ & $5 \times 10^9 M_{\odot}$  \\\hline
        \vspace{0.2cm} \\ \hline\hline
         $M_{G,max}^{Irr}$ & $M_{G,max}^{Spi}$ & $M_{G,max}^{Ell}$ \\ \hline 
        $5 \times 10^{9} M_{\odot}$ & $5 \times 10^{11} M_{\odot}$ &  --- \\ \hline
    \end{tabular}\\
         \vspace{0.2cm}
    \begin{tabular}{c c c c}
    \hline\hline
        $f_{ell}$ & $n^*$ & $\alpha$ & Mag$_V^*$\\ \hline
         0.74 & 52 & -1.15 & -21.30 \\ \hline
         \end{tabular}
    \caption{Fiducial model parameters for the single LF, consistent with the median found by \citet{moretti15} and \citet{mamon19} on the full WINGS cluster sample. The first and second rows are the lower ($M_{min}$) and upper ($M_{max}$) mass limits for irregular, spiral, and elliptical galaxies (Irr, Spi, Ell). On the bottom row, $f_{ell}$ is the fraction of galaxies that are elliptical/S0. $\alpha$ is the slope of the single LF. $n^*$ is the richness. Mag$V^*$ is the V-band magnitude identifying the break luminosity.}
    \label{tab:singleLF}
\end{table}

\begin{table}
    \centering
    \begin{tabular}{c c c c}
    \hline\hline
        $\alpha_b$ & $\alpha_f$ & Mag$_{V,b}^*$ & Mag$_{V,f}^*$\\ \hline
         -0.97 & -0.6 & -21.15 & -16.30 \\\hline
         \end{tabular}\\
    \caption{Fiducial model parameters for the double LF, consistent with the median found by \citet{moretti15} on the full WINGS cluster sample. The subscripts $_b$ and $_f$ represent the bright and faint end of the double LF respectively.}
    \label{tab:doubleLF}
\end{table}

For simplicity, we assume that all morphological types start evolving at the same time at high redshift ($z = 5$).  { We remind that with the calculations presented above we obtain a monolithic evolution, with galaxies of different morphological type that evolve separately. The inclusion of the hierarchical scenario is presented next.}

\subsubsection{Hierarchical scenario}\label{sec:hierarchical}
We test hierarchical scenarios of galaxy cluster formation by employing the technique presented in \citet{vincoletto12} and already coupled to chemical and dust evolution models in \citet{gioannini17b}. Instead of assuming that all moprhological types are born simultaneously and evolve separately, we let the number density of galaxies of morphological type $X$ evolve with redshift:

\begin{equation}
    n_X(z) = n_{X,0} (1 + z)^{\theta_{X}},
\end{equation}

\noindent where $n_{X,0}$ is the number density at redshift $z = 0$ for a morphological type $X$; $\theta_{X}$ is a parameter that for irregular, spiral, and elliptical galaxies is calibrated in \citet{gioannini17b} to be respectively 0.0, 0.9, and -2.5. From this number density, we derive the rescaled elliptical fraction:

\begin{equation}
	f_{ell} = f_{0}\frac{ (1+z)^{-2.5}}{f_{0} (1+z)^{-2.5} + (1 - f_{0}) (1+z)^{0.9}}, 
\end{equation}

\noindent { where $f_{0}$ is the elliptical fraction at present time. Notice that, if we derive similarly the spiral fraction $f_{spi}$, the relation $f_{spi} = 1 - f_{ell}$ is preserved across cosmic history.

Even for this scenario, we assume $z=5$ as our starting time for galaxy evolution.}

\subsubsection{Scaling of the LF with radius} \label{sec:modelradius}
The calculation presented in Section \ref{sec:modelingbasic} is valid for a Schechter function contained within $R_{200}$. The LF fits provided in \citet{moretti15} are valid within $0.5 R_{200}$; fortunately, the shape of the Schechter function does not vary from  from $0.5 R_{200}$ to $R_{200}$ \citep{annunziatella17}. However, { in order to obtain a fair comparison with observations \citepalias[i.e.,][]{43planck16},} we will need to calculate the integrated dust mass at radii larger than $R_{200}$. { Specifically, we are interested in radii of 15 arcmin}.  { We apply this method only at redsfhits $z \gtrsim 0.1$, where $R_{200} \sim 15'$; in fact} we are not interested in a profile at smaller radii than $R_{200}$ as our goal is to compute the total dust cluster mass. 

In order to rescale our integral to larger radii we take advantage of the NFW model \citep{navarro96}. The mass contained within a radius $t = r/r_s$ is given by:

\begin{equation}
M_{scaled} = M_{vir} \frac{\ln{(1 + t)} - t/ (1 + t)}{\ln{(1+c)} - c / (1 + c)},
\end{equation}

\noindent { where $r_s \rightarrow 15'$ is the scale radius that will rescale our integration to the 15 arcmin observed by \citetalias{43planck16}. The scale radius is redshift-dependent}. 
We assume that the virial mass $M_{vir}$ is roughly represented by $M_{200}$.
{ The concentration $c$ varies depending on the morphology: elliptical galaxies are well described with a concentration of $c \sim 4$, and spiral galaxies} with $c = 0.85$ \citep[see Table 2 ][]{cava17}. 

\subsection{Chemical evolution models and dust prescriptions} \label{sec:dustmodels}

To reproduce the dust mass $M_{d}$ in the three morphologies, we adopt detailed chemical evolution models for galaxies including dust evolution \citep[see ][for further details]{gioannini17a}. These models are built for different morphological types \citep[e.g., ][]{vladilo18, palla19, palla19b}. In this scheme, we assume that galaxies form by an exponential infall of gas on a preexisting dark matter halo: the evolution of an element $k$ within a galaxy then takes the following form:

\begin{equation}
    \label{eq:chem_evo}
    \dot{G}_k(t) = -\psi(t)\, X_k(t) + R_k(t) + \dot{G}_{k,inf}(t) - \dot{G}_{k,w}(t),
\end{equation}

\noindent where $G_k(t) = G(t)\,X_k(t)$ is the mass of the element $k$ in the ISM, normalized to the total mass; $X_k(t)$ is the fraction of the $k$-th element at time $t$. $\psi(t)$ is the SFR normalized to the total infall mass $M_G$:  { we adopt for it the Schmidt-Kennicutt law \citep{schmidt59, kennicutt89} $\psi(t) = \nu G(t)^k / M_{G}$ with $k = 1$. $G(t) = M_{ISM}(t)/M_{G}$ traces the evolution of the ISM mass normalized to the infall mass. $\nu$ is the star formation efficiency (expressed in Gyr$^{-1}$, which varies depending on the morphology.  We take $\nu = $ 15, 1, and 0.1 for elliptical, spiral and irregular galaxies, respectively.} $R_k(t)$ represents the returned fraction of an element $k$ that a star ejects into the ISM through stellar winds and supernova (SN) explosions, whereas $\dot{G}_{k,inf}$ and $\dot{G}_{k,w}$ account for the infall of gas and for galactic winds, respectively. { The mass outflow rate of an element $k$ due to galactic winds is defined as $\dot{G}_{k, w} = w_k \psi(t)$. $w_k$ stands for the mass loading factor for an element $k$, which is the same for every chemical species. We adopted a galaxy SF history that reproduces the majority of observable properties of local galaxies. In particular, the present time SFR, the SN rates and the chemical abundances  \citep{grieco12, gioannini17b}. 
}

Models also follow the comprehensive processes that influence dust evolution. For a specific element $k$ in the dust phase, we have:

\begin{equation}
    \label{eq:dust_evo}
\begin{split}
\dot{G}_{k,dust}= - \psi(t)\, X_{k,dust}(t) + \delta_k\, R_{k}(t) + {G}_{k,dust}(t)/\tau_{k,accr} +\\
-{G}_{k,dust}(t)/\tau_{k,destr}  - \dot{G}_{k,dust,w}(t), \hspace*{1.7cm}
\end{split}
\end{equation}

\noindent where $G_{k,dust}$ and $X_{k,dust}$ are the same of Equation \eqref{eq:chem_evo}, but for only the dust phase. This last equation includes dust production from AGB stars and core-collapse SNe ($\delta_{k}\,R_{k}$), accretion in molecular clouds (${G}_{k,dust}/\tau_{k,accr}$), and destruction by SNe shocks (${G}_{k,dust}/\tau_{k,destr}$). { $\dot{G}_{k, dust, w}(t)$ represents the mass outflow rate of the dust element $k$ due to galactic winds.}
 
\begin{figure*}
    \centering
    \includegraphics[width = \columnwidth]{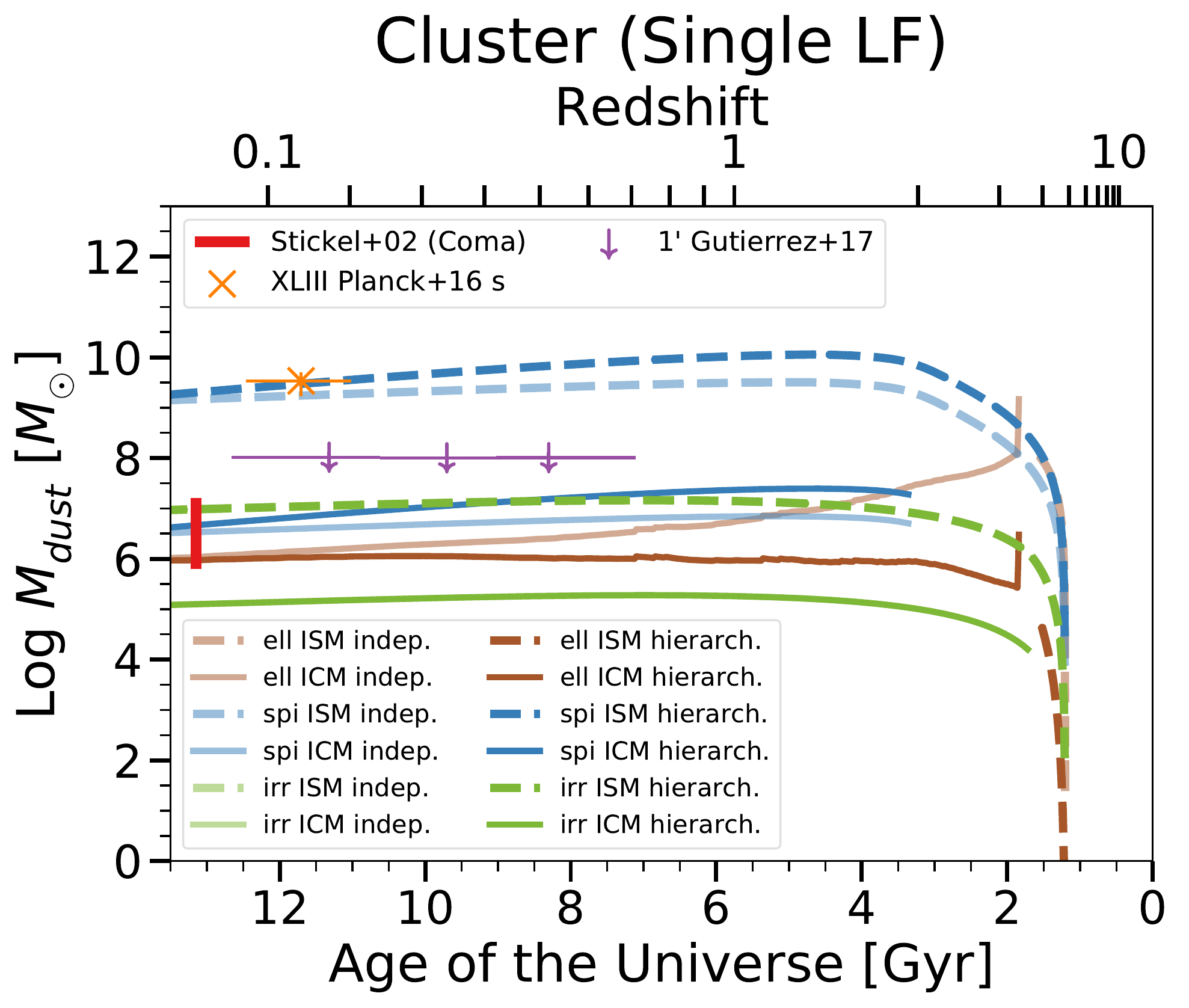}~
    \includegraphics[width = \columnwidth]{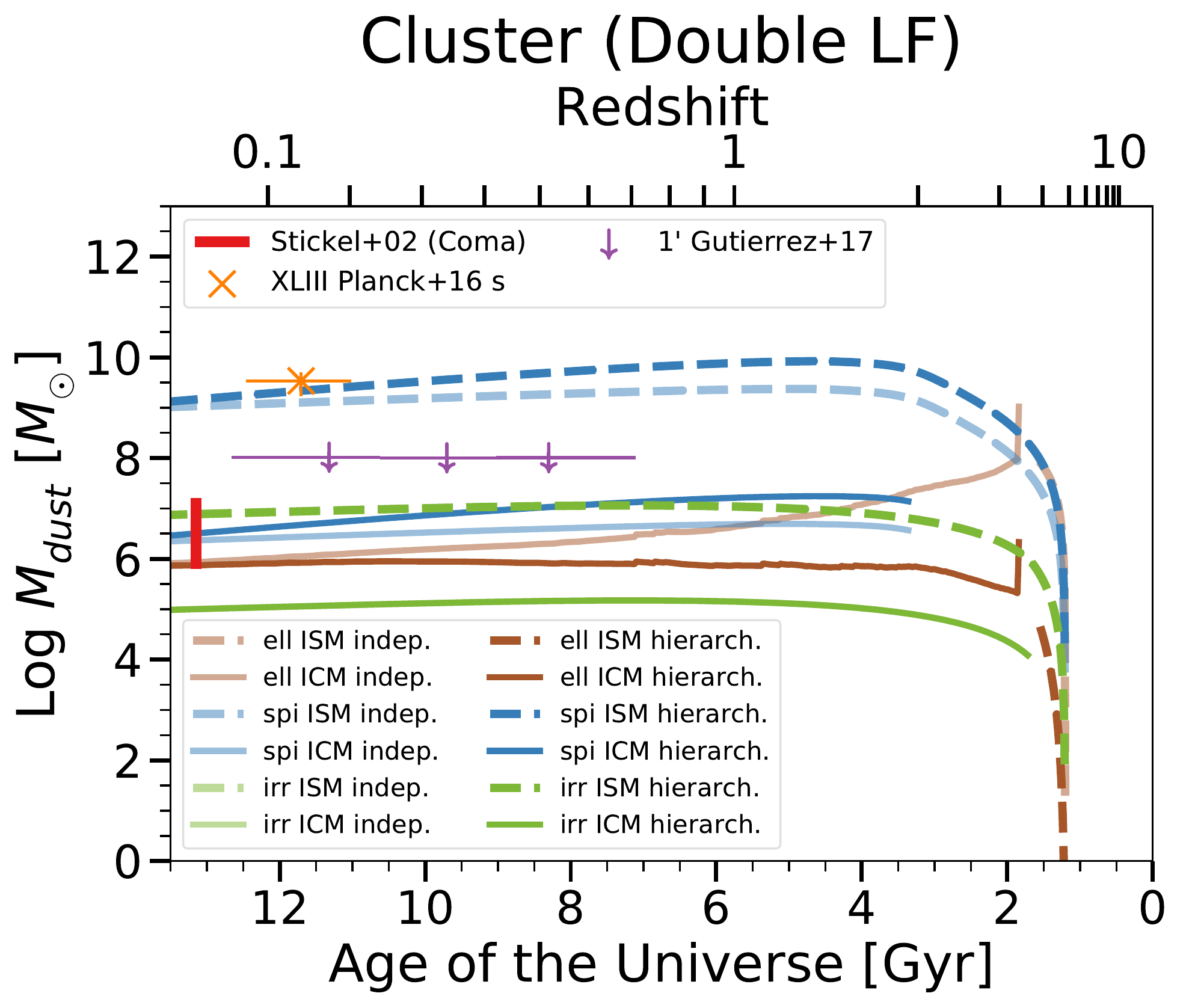}
    \caption{{ The evolution of dust mass components within a WINGS-like average cluster ($M_{200} \sim 5 \times 10^{14}M_{\odot}$, $f_{ell} = 0.74$, $n^* = 52$). The brown, blue, and green lines represent elliptical/S0, spiral, and dwarf irregular galaxies respectively. The dashed lines identify the integrated dust component residing within galaxies (ISM), whereas the solid lines identify the ejected components (ICM). The semi-transparent lines are computed assuming that each morphology evolves independently, whereas the fully opaque lines are computed assuming a hierarchical clustering scenario. All morphologies} begin their evolution 1.2 Gyr after the Big Bang (or at redshift $z \sim 5$). The orange '$\times$' is the \citetalias{43planck16} subsample at low redshift, whose average cluster mass is $M_{200} = 4.3 \times 10^{14} M_{\odot}$. The purple downward arrows are { the upper limits found in} \citetalias{gutierrez17} for the 350 $\mu$m \emph{Herschel} beam within a radius of 1 arcmin. The three redshift bins are centered around redshift $z = 0.173$, 0.338, and 0.517; they include clusters with total masses $> 10^{14} M_{\odot}$. In red is the dust estimate from \citet{stickel02} for the Coma cluster.}
    \label{fig:clus}
\end{figure*}

To compute the terms, we adopt detailed prescriptions from  literature. For the condensation efficiencies $\delta_k$ -- i.e. the fraction of an element expelled by stars in the dust phase -- we use prescriptions reported by \citet{piovan11}, whereas for the processes of accretion and destruction we adopt the metallicity-dependent timescales $\tau_{k,accr}$ and $\tau_{k,destr}$ from \citet{asano13}. We assume dust production by Type Ia SNe. This mechanism is a dubious contributor to the galactic dust budget. SN Ia will  produce dust locally and over short timescales -- \citealp{gomez12}, in fact,  observed large dust masses around the centenary Kepler and Tycho Type Ia SNe. However, \citet{nozawa11} computed that nearly no SNe Ia-born dust will survive the SN feedback -- at least, not long enough to be injected in the ISM. For spiral and irregular galaxies, the inclusion of SN Ia-born dust hardly affects their dust evolution profiles. Elliptical/S0 galaxies will instead be heavily affected; in fact, this morphology will otherwise host only low-mass short-lived AGBs that will produce little dust mass, and only sporadically.  In the case of Type Ia SNe, we adopt the prescriptions by \citet{dwek98} as implemented in \citet{calura08}: a dust condensation efficiency of 0.5 is taken for C, whether a value of 0.8 is taken for Si, Mg, Fe, Si, Ca, and Al. It is then assumed that for each atom of these 6 elements, one oxygen atom is also condensed, i.e., $M_O = 16 \sum_i 0.8 M_i / \mu_i$, where $M_i$ is the mass and $\mu_i$ is the atomic weight of each of the 6 elements $i$. While there could be other mechanisms of dust production within elliptical/S0 galaxies, such as dust growth in shielded shocked gas e.g. in AGB winds \citep{li19}, these processes cannot be easily included in our models; a rough estimate shows that their contribution to dust masses in ellipticals would be smaller than what would be produced by Type Ia SNe. We therefore implement SN Ia dust production both to test this model itself, and as an upper limit to what elliptical galaxies would be capable of contributing to the total cluster budget.

We then apply thermal sputtering to the dust component ejected from galaxies into the ICM, as prescribed in \citet{tsai95}. Assuming a fixed grain size of $0.1 \mu$m, the initial sputtering timescale is taken to be $\tau_{sp} = 5.5 \times 10^7$ yr, as derived by \citet{gjergo18}.
Specifically, the dust mass differential varies as $\dot{M}_{dust} = M_{dust} / \tau_{sp}$.  { This treatment of sputtering is appropriate for virialized clusters whose ICM has become hot, diffuse, and highly ionized, therefore it applies to redshifts as far back as $z \sim 1$ to 2 (i.e., the observational data we consider are characterized by a heavily sputtered ICM).}

\section{Results}\label{sec:results}
In this section we present the results of our method. In Section \ref{sec:fig1} we show the dust mass evolution for every cluster component: the ISM and ICM dust masses of elliptical, spiral, and irregular galaxies -- in the case of independent or hierarchical evolution -- when applied to single or double LFs. In Section \ref{sec:fig2} we rescale our total dust evolution to a fixed aperture of 15 arcmin, which is what \citetalias{43planck16} computes, and we normalize the curve to the cluster's $M_{200}$. We consider the total evolution of single and double LF for both the independent and the hierarchical scenario.  In said plot, we also vary one of the most important parameters of our model: $f_{0,ell}$. In Section \ref{sec:fig3} we finally present a variation of 7 of the main single LF function parameters for the 6 components -- elliptical, spiral, and irregular, for ISM and ICM -- with both the independent and the hierarchical scenarios. 

\subsection{Component evolution of a characteristic cluster} \label{sec:fig1}
{ Figure \ref{fig:clus} presents the results of our integration method within $R_{200}$ of a  cluster of average LF parameters \citep{moretti15} and average cluster parameters \citep{mamon19} of the WINGS cluster sample, as specified in Table \ref{tab:singleLF} and Table \ref{tab:doubleLF}. On the right, we employed a single LF; on the left, a double LF. The dust mass components
coming from the elliptical/S0, spiral and dwarf irregular galaxy morphologies are represented respectively with colors brown, blue and green. The two line styles, dashed and solid, identify the ISM dust components residing within galaxies and the ICM dust component ejected in the hot intergalactic and intracluster media, respectively. 
The fully opaque curves follow a hierarchical clustering scenario. The faint spiral and elliptical curves trace the scenario in which the morphologies evolve independently. Given that for irregular galaxies $\theta_{irr} = 0$, the dust mass does not change between scenarios. For elliptical galaxies, our ISM component is instantaneously ejected into the ICM as soon as stellar winds ignite.}
The dominant component for both galaxy evolution scenarios and both LFs is the ISM dust in spirals, in agreement with the interpretation proposed by \citet{roncarelli10} that late-type galaxies should dominate the overall dust IR emission in galaxy clusters. We confirm this conclusion through our optically-calibrated dust evolution model. For both scenarios and for both LFs, the next most abundant dust mass component at a redshift $z < 2$ appears only around 2 dex below the ISM spiral component.

{ As expected, the choice of single or double LF does not affect significantly the most massive galaxies, if not by a marginal decrease in dust mass. More surprisingly, it does not seem to affect the  dwarf irregular components. Regarding these components, we see that below $z \lesssim 0.5$ the dust mass within irregular galaxies exceeds the dust mass ejected in the ICM by spirals.} 

There is a minimal but steady dust mass loss within the ISM of spiral galaxies in both hierarchical and monolithic evolution scenarios after the peak at a redshift of $z \sim 2$, at a galaxy age of $\sim 2$ Gyr. The slope traced by the spiral ICM dust line is steeper in the hierarchical scenario due to the higher spiral fraction at earlier times. The trend is flipped for elliptical ICM dust, as this morphology was less numerous in the past for the hierarchical evolution.

The observational data included in the Figure are: the low-redshift total cluster dust mass by \citetalias{43planck16} bin with average cluster mass of $\left< M_{200} \right> = 4.3 \times 10^{14} M_{\odot}$ and average redshift $\left< z \right> =  0.139$ (orange cross); the ICM dust mass upper limit estimated by \citetalias{gutierrez17} in the $350 \mu$m channel 
within 1 arcmin (purple downward arrows), applied to the three redshift bins of the cluster sample with virial masses $> 10^{14} M_{\odot}$; the \citet{stickel02} estimate for the Coma cluster (red vertical dash). This latest value, while near the cirrus foreground noise \citep{kitayama09}, is a good upper limit for { ICM dust of local quiescent clusters. Many of our curves of the cumulative ICM dust components fall within the \citet{stickel02} ICM dust mass estimate for galaxy evolution scenarios and LFs.} 
{ Our results are also in agreement with the dust ICM upper limit estimate by \citetalias{gutierrez17} and with total cluster dust found by \citetalias{43planck16}.}

Summarising, within $\sim 2$ Gyr since galactic birth, the overall ISM of cluster galaxies contains already about $10^{10} M_{\odot}$ in dust mass for a massive WINGS-like cluster ($3 \times 10^{10} M_{\odot}$ in dust mass for the independent morphological evolution scenario). { This corresponds to a dust-to-gas ratio of $DtG \sim 5 \times 10^{-4}$ for the entire cluster. The ICM components are between 2 to 4 dex lower, around or below a dust mass of $10^7 M_{\odot}$, which corresponds to a low DtG of the order of $10^{-6}$.}

\begin{figure}
    \centering
    \includegraphics[width = \columnwidth]{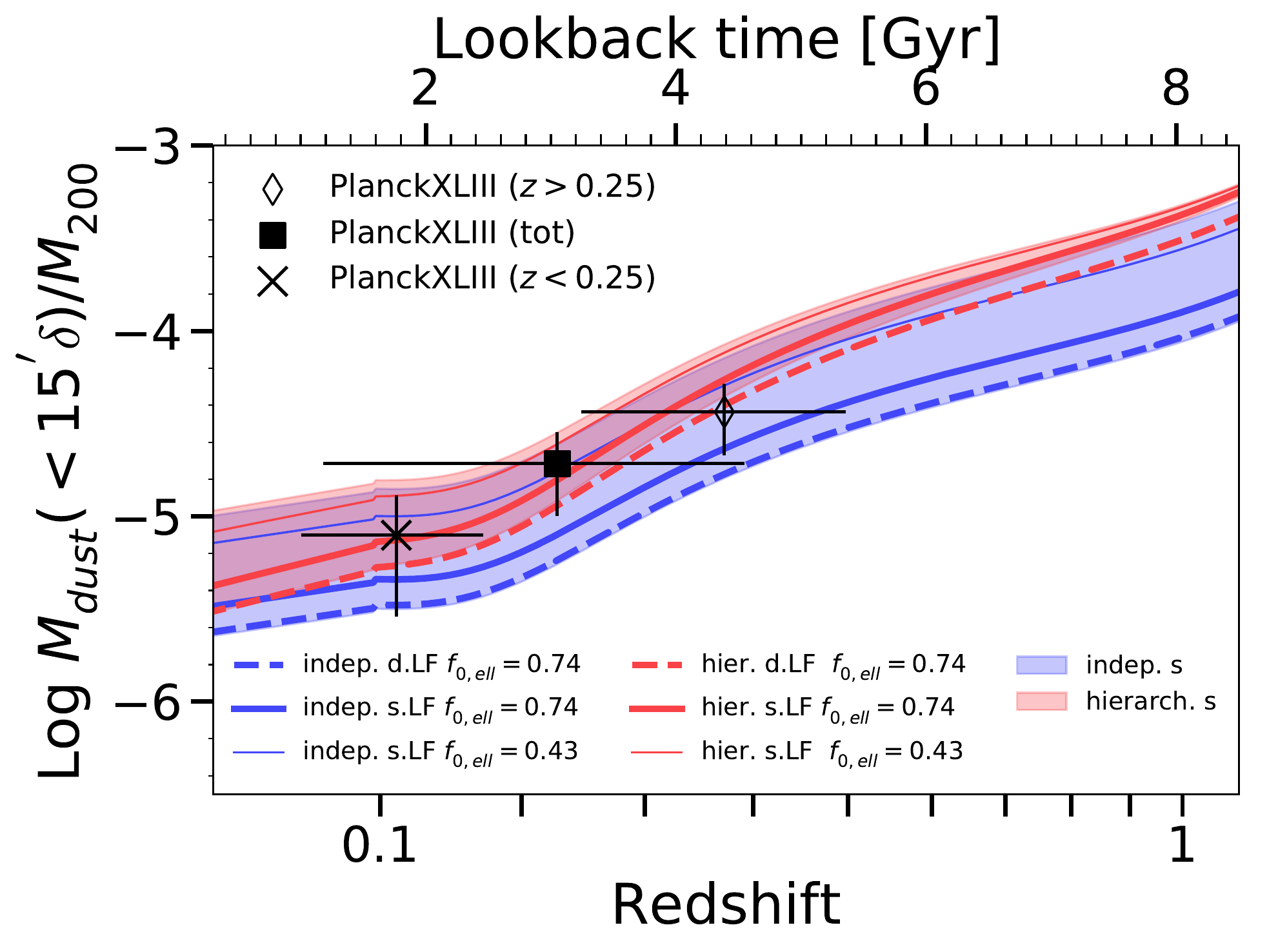}
    \caption{{ The black points represent the average Planck dust mass measured within 15 arcmin of a stacked cluster sample \citepalias{43planck16}.  All 3 masses are normalized to the average $M_{200}$ of their respective bins.} The full square is the full sample ($M_{200} \simeq 5.6 \times 10^{14} M_{\odot}$ and $z \sim 0.26$). The '$\times$' is the nearby ($z < 0.25$) bin of average mass $M_{200} \simeq 4.3 \times 10^{14} M_{\odot}$, the hollow diamond is the distant ($z > 0.25$) bin ($M_{200} \simeq 7.0 \times 10^{14} M_{\odot}$).  { The curves follow the integrated dust evolution computed in the present work, as they would be seen within a fixed aperture of 15 arcmin from the cluster center, normalized by $M_{200}$. We perform this rescaling to mimic the Planck data. The blue curves identify the case in which the morphologies evolve independently, whereas the red curves show the case in which morphologies merge in accordance to  hierarchical clustering. The solid curves assume a single LF, and the dashed curve a double LF, calibrated on the WINGS cluster survey \citep{moretti15}. The thick curves assume an elliptical fraction of $f_{0,ell} = 0.74$, as in the WINGS clusters. The fainter curves assume instead a $f_{0, ell} = 0.43$. This is the fraction of the Virgo cluster and it is representative of other similarly young and active clusters. The semi-transparent red and blue bands span -- for the independent and hierarchical evolution -- an $f_{0,ell}$ between 0.2 and 0.82, the latter being the elliptical fraction of the Coma cluster, representative of fully evolved clusters.}}
    \label{fig:15arcmin}
\end{figure}

\begin{figure*}
    \centering
    \includegraphics[width = \linewidth]{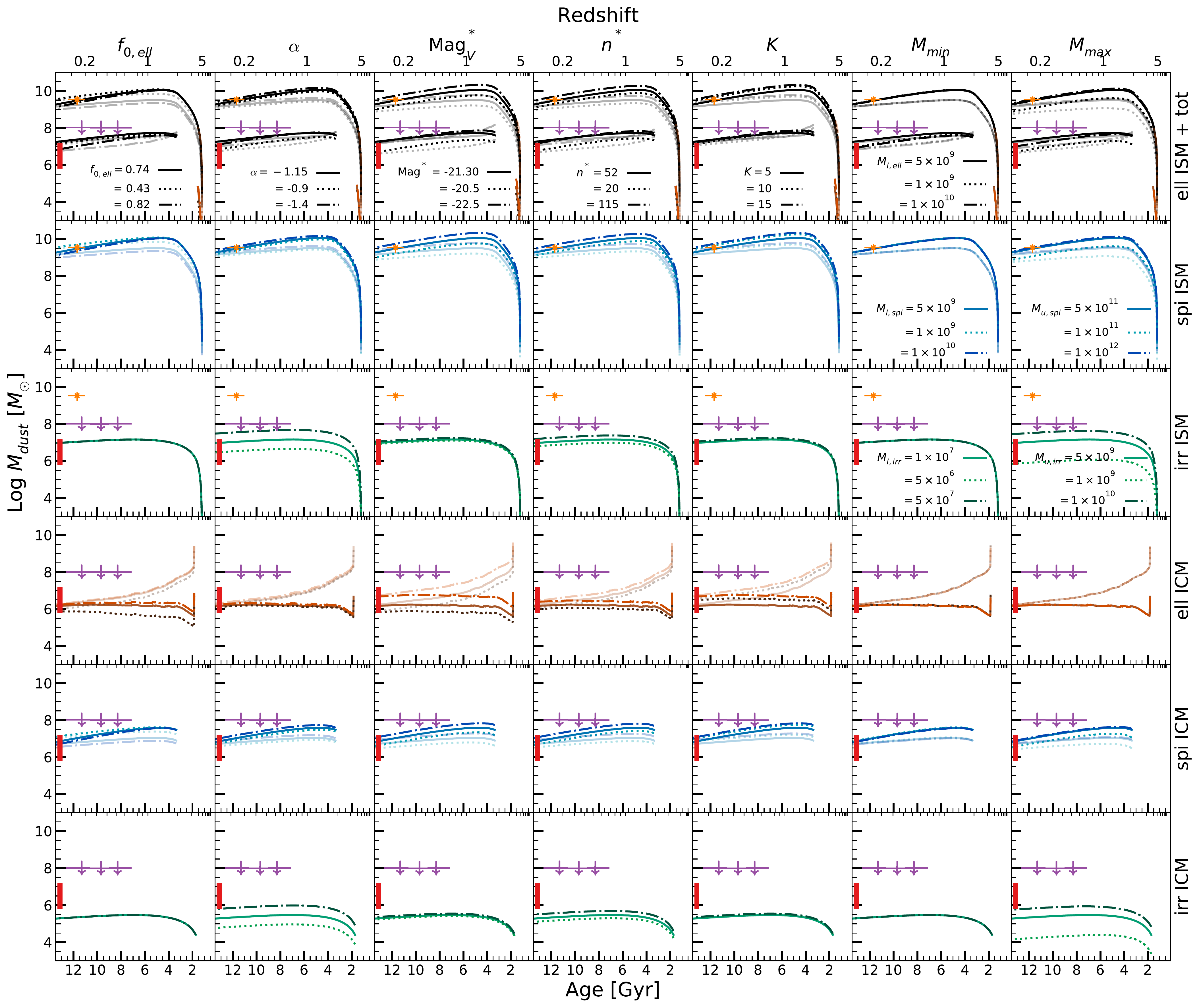}
    \caption{{ In this plot we show for each morphology (rows) the evolution of the individual dust mass ISM components (the colored curves in the first 3 rows) and ICM components (the bottom 3 rows) and their dependence on our model parameters (columns). Elliptical galaxies are shown in orange/yellow, spiral galaxies in blue, and irregular galaxies in green. The first row, aside from the orange/yellow ISM elliptical component, contains black lines that display the total ISM evolution, and total dust ICM evolution with the addition of the irregular ISM. We integrate only over the single LF. The faint lines in rows 1, 2, 4, and 5 trace the independent morphological evolution, whereas the fully-opaque lines trace the hierarchical evolution. Notice that, given that $\theta_{irr} = 0$ in our model (Section \ref{sec:hierarchical}), there is no difference between hierarchical and independent evolution for irregular galaxies. All fiducial curves are shown with a solid line. From left to right, in each column we vary the final fraction of elliptical galaxies $f_{0,ell}$, the slope of the LF $\alpha$, the V-band magnitude at the break of the LF Mag$^*_V$, the cluster richness $n^*$, the mass-to-light ratio $K$, and the lower and upper infall mass integration limits $M_{G,min}$ and $M_{G,max}$. For the first 5 columns, the parameter variation is labeled in the first row. The parameter variation is identified as specified in said labels with dotted and dash-dotted lines. For the last two columns, the integration limits of elliptical, spiral, and irregular galaxies are shown respectively in the first, second, and third row. The choice of parameter variation reflects standard values from literature, as explained in the text. Our model falls tightly within the observational constraints. The data points from \citetalias{43planck16, gutierrez17}, and \citet{stickel02} are identical to Figure \ref{fig:clus}. In the case of \citetalias{gutierrez17}, we have omitted the dust mass upper limits within 5 arcmin as they extend beyond the $\sim R_{200}$ cluster radii we trace in our evolution. We show their dust mass upper limit within 1 arcmin. }}
    \label{fig:variations}
\end{figure*}

\subsection{Mimicking Planck's observations }  \label{sec:fig2}

Figure \ref{fig:15arcmin} rescales our model curves so that they mimic \citetalias{43planck16}, from which the three data points are taken. The middle full square is the dust mass estimate from the whole sample of 645 stacked clusters in redshift range $0.01< z < 1.00$ (with a $M_{200}$ of $5.6 \pm 2.1 \times 10^{14} M_{\odot}$). The other two points are the redshift bins of two subsamples at $z \leq 0.25$ with 307 clusters { ('X', with a $M_{200}$ of $4.3 \pm 1.7 \times 10^{14} M_{\odot}$), and at $z > 0.25$ with 254 clusters (hollow diamond, with a $M_{200}$ of $7.0 \pm 1.5 \times 10^{14} M_{\odot}$)}.  As the dust masses in \citetalias{43planck16} are integrated out to radii of 15 arcmin, we rescale our dust evolution curves out to the same volumes above $z \sim 0.1$, whereas at lower redshift we adopt an unscaled profile (see Section \ref{sec:modelradius}). 
Both data and curves are normalized to their respective $M_{200}$. In the theoretical curves, the $R_{200}(t)$ rescaling to 15 arcmin and $M_{200}(t)$ are taken from the D2 run in \citet{gjergo18}. 

The blue curve shows the case in which the morphologies evolve independently (according to the monolithic scenario), the red curve shows instead the hierarchical clustering iteration. The solid lines are computed using the single LF, the dashed lines represent instead the double LF. We adopt, just like in the previous case, LF and cluster parameters coming from the WINGS average. The WINGS elliptical fraction $f_{0,ell} = 0.74$ \citep{mamon19} is shown with bold lines. The elliptical fraction is the parameter we vary in the other curves.  The thin solid curves assume a final elliptical fraction of $f_{0,ell} = 0.43$. This is the fraction observed in the dynamically young local Virgo cluster. The two semi-transparent bands, red for the hierarchical scenario and blue for the independent morphological evolution, span an elliptical fraction from 0.2 on top to 0.82 on the bottom for the single LF. We chose $f_{0,ell} = 0.82$ as our lower limit for this band as this is the elliptical fraction observed in the mature Coma cluster \citep[][]{melnick77}. We see that varying the elliptical fraction in the range between $\sim 0.2$ to $\sim 0.8$ changes the dust mass estimates for each model by not more than a factor of 6 overall. In the range encompassed by the Planck parameters, the slope is mainly driven by the larger fixed aperture. However, by $z \sim 0.5$ to 0.6, the angular size has already reached 3 Mpc, and the NFW rescaling doesn't increase the dust mass by much, due to the low galaxy density out to these outskirts. The increase at higher z is mainly caused by the normalization to $M_{200}$. 

The data-model comparison seems to weakly favor a hierarchical scenario with the WINGS average elliptical fraction of 0.74, with no significant preference between the single or double LF. For $f_{ell} = 0.74$, $\chi^2$ is respectively 1.10 and 0.40 for the double and single LF. For the single LF with $f_{ell} = 0.43$, $\chi^2$ is 0.16. For the hierarchical scenario, the $\chi^2$ in the curves with $f_{ell} = 0.74$ is 0.07 for the double LF, and 0.21 for the single LF; 0.80 is the value of the single LF with the $f_{ell} = 0.43$. As $f_{ell}$ (and so also $f_{spi}$) is time-dependent in the hierarchical scenario, at higher redshifts the red curves converge to higher dust-to-M$_{200}$ ratios. The difference between the constant elliptical fraction of the monolithic evolution and the same fraction for the hierarchical evolution widens at high final elliptical fraction values because the spiral number density falls more steeply (and the elliptical number density grows more rapidly). 
 Interestingly, our curves fall right between the data without the need for any calibration. This is the first theoretical work that reproduces the Planck total cluster dust mass estimates at low-to-intermediate redshift, without amending the sputtering timescales or other model parameters.

\subsection{Parameter variations}  \label{sec:fig3}
{ Figure \ref{fig:variations} displays the dependence of our model on the single LF parameters, which from left to right column are: the elliptical fraction $f_{0,ell}$, the LF slope $\alpha$, the V-band magnitude at the break of the LF Mag$^*_V$, the cluster richness $n^*$, the mass-to-light ratio in $K$, the lower $M_{l,X}$ and upper $M_{u, X}$ mass integration limits for each of the three morphologies $X$. On each row we display the dust mass components: the top three rows contain the ISM dust of elliptical (orange/yellow), spiral (blue), and irregular galaxies  (green), and the bottom three rows contain the ICM components, with same colors and order. The first row also contains in black the total ISM components and the total ICM components. Faint curves trace the independent morphological evolution scenario, whereas the fully-opaque curves the hierarchical scenario. Fiducial  values are plotted with the solid line, and are identical to Figure \ref{fig:clus}, as are the observational data points. The variations, as identified by the label, are shown with dotted and dot-dashed lines. 
} 

Each parameter encompasses the typical average ranges of known clusters. $f_{0,ell} $ spans from 0.43 to 0.82, which is the elliptical fraction of the Virgo Cluster and of the Coma cluster, respectively. 
$\alpha$ and Mag$^*_V$ test ranges which include $\sim 1.5 \sigma$ of the LF slope and break V-band magnitude in \citet{moretti15}. Specifically, $\alpha$ ranges between -0.9 and -1.4, and Mag$^*_V$ ranges between -20.5 and -22.5. For the richness, we once again refer to the WINGS sample: we vary $n^*$ between 115 \citep[highest richness considered in ][]{schechter76} and 52 (average WINGS richness). The $K$ spanned range is between 5 and 15: typical $K$ values in early-type galaxies range from  $\sim$ 5 to $\sim$ 13, while in late-type galaxies it ranges between 5 and 10 \citep{demasi19}. For galaxies of equal mass, $K = 10$ will mean galaxies are half as luminous as those at $K = 5$. 

In our cluster dust mass estimates, the model depends linearly only on $f_{0,ell}$ and on $n^*$. 
The richer and younger the cluster, the more dust there is, but the cluster richness does not vary the cumulative dust mass if not by a factor of 3. A factor of 3 is also the extent to which the variation of $f_{0,ell}$ changes the dust mass for the spiral and elliptical components. Notice that the dust mass evolution with a Virgo-like elliptical fraction is slightly larger than the total dust Planck estimate at $z \sim 0.2$. 
In each case at low redshifts, the spiral ICM dust component is between 3 and 30 times larger than the elliptical ICM component. Thus, spiral galaxies are more efficient than elliptical/S0 galaxies at contaminating the ICM with dust. 

A little more complicated is the issue of the contribution by smaller galaxies. These faint irregular galaxies are difficult to resolve, so the emission and extinction from dust within these systems may be confused as newly ejected ICM dust. To compare how much dust there is within irregular galaxies with each parameter variation against the \citetalias{gutierrez17} ICM upper limit, we have added those data points to the third row of the plot. As expected, the irregular-born dust mass has little dependence on most parameter variations. The only difference, a significant one, is generated by the choice of the $\alpha$ LF slope, and of an upper mass integration limit $M_{max,irr}$ ($M_{u,irr}$ in the plot): in fact, these are the only cases where parameter changes in a certain scenario span more than a factor 6 (3 in most of the other cases). The increase in dust mass in the $\alpha$ variation is mostly due to the excess in number density on the more massive end of the integration limits, closer to the break magnitude. In fact, when we test these variations on the double LF function, we find that a steeper increase of $\alpha_f$ provides only a marginal gain in dust mass. That said, the upper bound for irregular galaxies is dictated by observations \citep[e.g.,][]{vaduvescu07, yin11}. A galaxy with an infall mass of  $10^{10} M_{\odot}$ will likely be a spiral rather than an irregular galaxy. Therefore, it is not realistic to extend the upper integration limit for irregular galaxies above our fiducial of $5 \times 10^9 M_{\odot}$.


Except for the above case of the irregular components, the lower and upper mass bounds of our integration matter little. Unless the infall mass is of the same order or larger than the break infall galaxy mass ($M^*_G = 6.94 \times 10^{10} M_{\odot}$ for the single LF), the integration is not very sensitive to either mass bounds. We see, in fact, that if we were to choose upper bounds close to $M^*_G$, the integration would not be stable for irregular and also spiral galaxies.  In the case of spiral galaxies, $M_{u,spi} = 10^{11} M_{\odot}$ -- an upper bound five times smaller than the fiducial -- decreases the dust mass by a factor of 3. The upper bound $M_{u,spi} = 10^{12} M_{\odot}$, five times larger than the fiducial upper bound, hardly makes a difference -- the dust mass has nearly reached its limiting value. 
A dependence on the upper limit does not exist for elliptical galaxies, because we integrate them over a single upper incomplete gamma function having only a lower bound. Even if we were to impose an upper limit, as long as it were not close to $M^*_G$ (e.g., $M_{u,ell} = 10^{12} M_{\odot}$) it would not affect the stability of our results.

In Figure \ref{fig:variations}, we do not plot the dependencies on double LF parameters. In fact, by testing all the other cases with the double LF, we find only minor differences.

Concluding, both in the case of the ISM and of the ICM components, spiral galaxies are dominant and sufficient to explain the observed dust abundances. After spiral galaxies, the ISM from irregular galaxies is a major dust source. This dust mass from irregulars comes mainly from the bright end of the single LF, as the dust mass is nearly independent of the lower integration limit. 

\section{Discussion and Conclusions} \label{sec:conclusions}

The latest and most advanced observations of dust in galaxy clusters come from the analysis of stacked signals in large cluster datasets. \citetalias{43planck16} provided a spatially-unresolved average total dust mass of over 600 clusters, using the dust thermal emission as detected by the Planck beams and IRAS data. \citetalias{gutierrez17} reported the ICM dust directly, using both emission and extinction techniques. In the case of emission, \citetalias{gutierrez17} subtracted foreground sources from the detected ICM emission; in the case of extinction, they estimated the ICM extinction curves around sources (quasars or galaxies) located on the background of galaxy clusters. 

With this work, we aimed at breaking down the galaxy-born total cluster dust mass into its various components. We differentiated the three main morphologies that contribute to the dust budget and we also separated dust residing within galaxies from the dust ejected in the ICM by applying sputtering to the dust ejected out of galaxies. 
Then we applied an integration method developed by \citetalias{matteucci88} to integrate galaxy dust over a parametrization of cluster LFs. Our LF parameters were chosen from medians and average values of large local galaxy clusters \citep[WINGs, ][]{moretti15, mamon19}. We tested the cluster dust evolution both in a monolithic (i.e., galaxies evolve as a whole and indipendently of other galaxies) and in a hierarchical scenario of galaxy formation \citep{vincoletto12}.

Our main conclusions are as follows:
\begin{itemize}
\item { Dust within spiral galaxies accounts for most of the dust contained within clusters in any assumed galaxy formation scenario}. 
We estimate that a typical cluster should have around $10^{10} M_{\odot}$ in total ICM dust mass, mainly residing within its spiral galaxies, or a DtG of $\sim 5 \times 10^{-4}$. Dust mass in the ICM of a cluster should be of the order of $10^7 M_{\odot}$.

\item { Dust ejected into the ICM by spirals ($\sim$2.5 dex less abundant than spiral ISM dust) is comparable in mass to the dust component residing within} irregulars, and it is largely consistent with the dust abundance upper limits measured in the ICM of local galaxy clusters \citep{stickel02, kitayama09, bai07} { regardless of the assumed spiral fraction}.

\item { 
Without any calibration
, we are able to produce dust abundances analogous to observational data \citepalias[i.e.,][]{43planck16, gutierrez17}. 
It is not necessary to rely on dust sources aside from those of galactic origin -- such as dust growth in cold filaments of the ICM, or dust generated around intracluster stars -- to explain where the bulk of cluster dust comes from. Galaxy-born dust is likely the main source of cluster dust. Unless the LF of such smaller sources were to spike at the faintest luminosities and were not Schechter-like, the smaller sources cannot produce sufficient dust mass to account for the large ICM dust upper limits reported by \citetalias{gutierrez17}. }


\item { Letting our model evolve according to a hierarchical scenario marginally improves the redshift-dependence of dust evolution, and better captures the $M_{dust}(< 15')/M_{200}$ slope observed by \citetalias{43planck16}.}

\item {Galaxies close to the break of the LF dominate  the cluster dust mass, and the method has little dependence on the integration limits. A similar result was found in \citet{gibson97} for the gas-phase metals in the ICM: galaxies at the break of the LF dominate the ICM metal enrichment. In the case of gas-phase metals, it is the contribution of the dominant early-type galaxies (elliptical/S0 in our model) to govern the enrichment, not spiral galaxies. However, ellipticals produce little dust in dust evolution models, so that they cannot contribute to the present time ICM dust mass as much as spirals do. 
 This conclusion holds even assuming, as we did throughout this work, a strong dust production by Type Ia SNe. Stellar winds in ellipticals are very efficient at driving out of the galaxy any newly-formed dust, so that any dust  observed  within ellipticals must be of recent origin; but also the ICM dust component coming from ellipticals will not survive for long due to sputtering.
 }

\item {By varying our model parameters ($f_{ell}$, $\alpha$, Mag$^*_V$, $n^*$, $K$, $M_{l,X}$, $M_{u,X}$) over a reasonable range chosen from literature, we find that dust masses change by up to a factor of $\sim $ 6 (up to 3 most of the times). The only exceptions are the slope of the LF, $\alpha$, and the upper mass limit of irregular galaxies on the irregular components only. All these variations do not affect deeply our conclusions, confirming the robustness of the method adopted.}

  \end{itemize}


\section*{Acknowledgements}
The authors thank Alessia Moretti for providing the single Schechter and double Schechter fits to the full WINGs cluster sample. A special thank you goes to the anonymous referee -- unofficial 7th author to this paper -- for the meticulous and  constructive feedback that improved significantly the quality of this manuscript.



\bibliographystyle{mnras}
\bibliography{gjergo_bibliography} 




\bsp	
\label{lastpage}
\end{document}